\documentclass[useAMS,usenatbib,usegraphicx]{mn2e}

\title[M31 star stream]{Tracing the star stream through M31
using planetary nebula kinematics}
\author[Merrett et al.]
{H.R.\ Merrett$^1$\thanks{E-mail: ppxhm@nottingham.ac.uk},
K.Kuijken$^{2,3}$,
M.R. Merrifield$^1$,
A.J. Romanowsky$^1$,
N.G Douglas$^3$,
\newauthor
N.R. Napolitano$^3$,
M. Arnaboldi$^4$,
M. Capaccioli$^4$,
K.C.Freeman$^5$,
O. Gerhard$^6$,
\newauthor
N.W. Evans$^7$, 
M.I. Wilkinson$^{7}$, 
C. Halliday$^{8}$, 
T.J. Bridges$^{9}$, 
D. Carter$^{10}$\\
$^1$School of Physics \& Astronomy, University of Nottingham, 
    University Park, Nottingham, NG7 2RD\\
$^2$Leiden Observatory, PO Box 9513, NL-2300 RA Leiden, The Netherlands\\
$^3$Kapteyn Institute, PO Box 800, NL-9700AV Groningen, The Netherlands\\
$^4$Osservatorio di Capodimonte, Via Moiariello 16, Naples 80131, Italy\\
$^5$Research School of Astronomy and Astrophysics, Australian National 
    University, Canberra ACT 2601, Australia\\
$^6$Astronomisches Institut, Universit\"at Basel, Venusstrasse 7, 
    CH 4102 Binningen, Switzerland\\
$^7$Institute of Astronomy, Madingley Road, Cambridge CB3 OHA, UK\\
$^{8}$Osservatorio Astronomico di Padova, Vicolo dell'Osservatorio 5, I-35122 
      Padova, Italy\\
$^{9}$Anglo-Australian Observatory, Epping, NSW 1710, Australia\\
$^{10}$Astrophysics Research Institute, Liverpool John Moores University, 
       Twelve Quays House, Egerton Wharf, Birkenhead CH41 1LD, UK 
}
\begin{document}

\date{Accepted 2003 ????? ??. Received 2003 ?????? ??; in original form 2003 September 29}

\pagerange{\pageref{firstpage}--\pageref{lastpage}} \pubyear{2002}

\maketitle

\label{firstpage}

\begin{abstract}
We present a possible orbit for the Southern Stream of stars in M31,
which connects it to the Northern Spur.  Support for this model comes
from the dynamics of planetary nebulae (PNe) in the disk of M31:
analysis of a new sample of 2611 PNe obtained using the Planetary
Nebula Spectrograph reveals $\sim 20$ objects whose kinematics are
inconsistent with the normal components of the galaxy, but which lie
at the right positions and velocities to connect the two photometric
features via this orbit.  The satellite galaxy M32 is coincident with
the stream both in position and velocity, adding weight to the
hypothesis that the stream comprises its tidal debris.
\end{abstract}

\begin{keywords}
Local Group -- galaxies: individual: M31 -- galaxies: interactions --
galaxies: kinematics and dynamics -- galaxies: structure
\end{keywords}

\section{Introduction}\label{sec:intro}
 It is now generally accepted that mergers play a key role in the
formation of galaxies \citep{wr78}.  Since galaxy evolution is an
ongoing process, we might therefore expect to catch a number of
systems in the nearby Universe mid-merger.  Indeed, dramatic major
mergers with complex tidal tails have been documented for quite some
time \citep[see, for example, ][]{s86}.  Perhaps of greater importance
to the more passive evolution of galaxies, and as a possible cause of
phenomena such as thick disks \citep{qhf93}, evidence for the more
common minor mergers is now coming to light.  In these cases, the
detritus of the events has a high enough surface brightness be visible
as a stellar stream.  The Sagittarius Dwarf Galaxy provides a fine
example in the Milky Way \citep{maj03}, and one need look no further
than the closest good-sized galaxy, M31, to find another dramatic
stream of stars \citep{ib01}, which is presumably the remnant of a
similar minor merger.  The proximity of these streams indicates quite
how common this phenomenon must be, but it also offers prime
laboratories for studying the merger process in detail.  The
Sagittarius Dwarf is inconveniently placed behind the centre of the
Milky Way, and also presents the usual problems of geometry when
trying to study an object inside our own galaxy, so the M31 stream is
probably the best candidate for analysis.  However, even this is not
without its problems: the large extent of M31 necessitates a very
extensive data set for any complete survey, and the low surface
brightness of the feature renders the stream hard to detect
particularly against any of the brighter parts of M31.  The faintness
of the stream and its constituent stars also makes it very challenging
to obtain the kinematic observations that would tie down the full
dynamical structure of the stream, and thus unequivocally demonstrate
its nature.

In this paper, we seek to overcome these difficulties by supplementing
the existing photometric data with a new sample of 2611 planetary
nebulae in the disk of M31.  The kinematics of these discrete
stellar tracers can be used to pick out the star stream right through
the brighter parts of M31's disk as well as helping determine its
dynamics.  The remainder of this paper is laid out as follows.  In
Section~\ref{sec:photo} we review the existing photometric data on the
M31 stream features and the hypothesis as to how they may be
connected.  Section~\ref{sec:kin} presents the kinematic data,
which strengthens the case for such a link.  Section~\ref{sec:orbit}
shows an orbit model that quantitatively marries these photometric and
kinematic features, and presents the evidence that the stream may
arise from M32's tidal debris.  Section~\ref{sec:conc} concludes.

\begin{figure*}
\includegraphics[width=177mm]{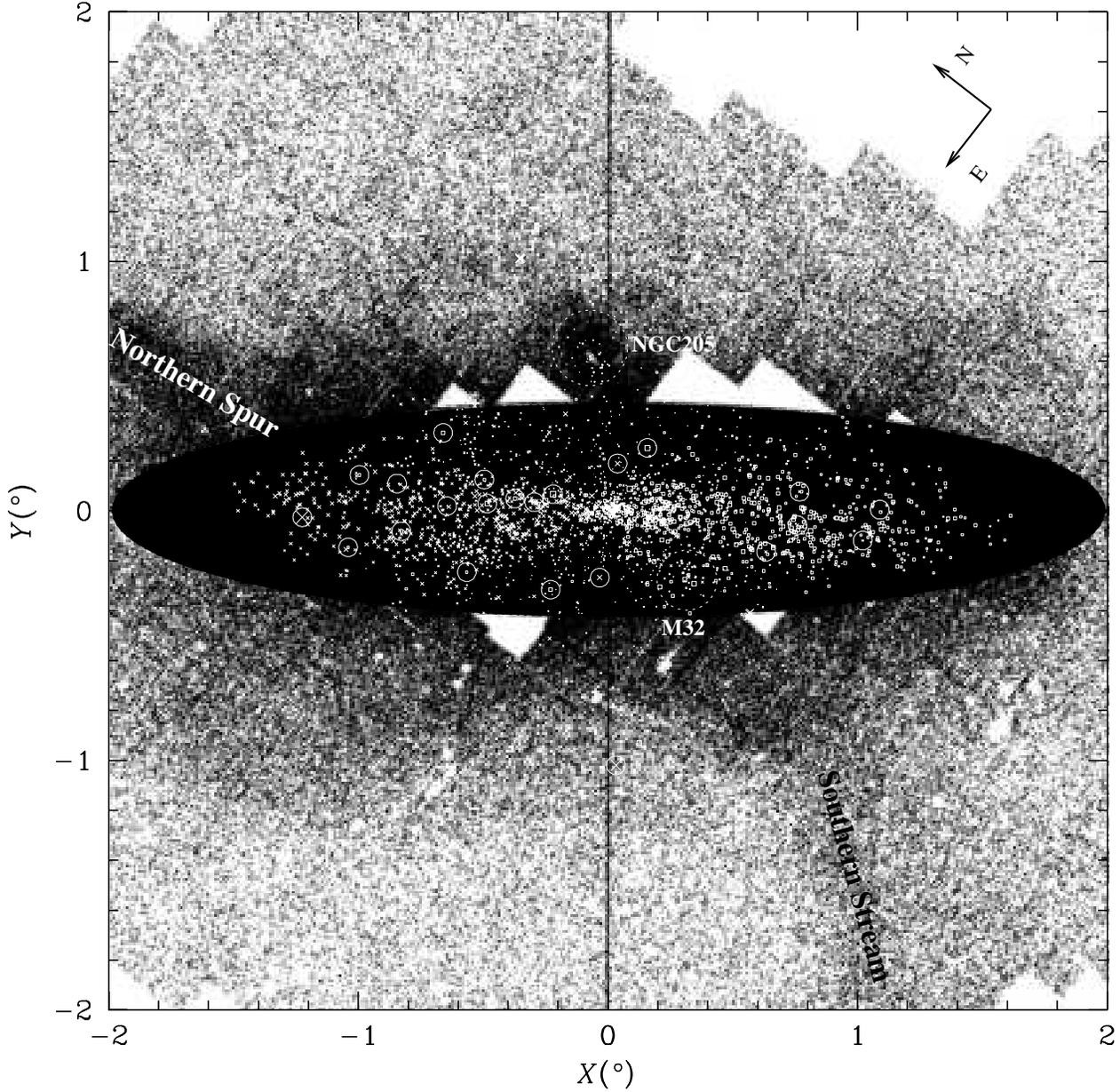}
\caption{Composite illustration of the data on the star stream in M31,
presented in a coordinate system in which $X$ lies along the major
axis of M31, increasing toward the SW, $Y$ increases along the minor
axis toward the NW, and $Z$ increases along the line of sight.  The
greyscale shows the star-count data, as published in Ferguson et al.\
(2002), with the Southern Stream and Northern Spur annotated.  The
black ellipse indicates the extent and inclination of M31's disk, and
the locations of NGC205 and M32 are also marked.  The points show the
PNe detected in the PN.S survey, with the size of the symbols encoding
their line-of-sight velocities relative to M31 (crosses receding,
boxes approaching).  The circled points are ``friendless'' PNe that
lie more than $4\sigma$ in velocity from their 30 nearest neighbours.}
\label{fig:data}
\end{figure*}

\section{The photometry of the stream}\label{sec:photo}
The star stream in M31 was first reported by \citet{ib01}, who
demonstrated that this dramatic low-surface-brightness linear feature
protrudes to the south-east of the galaxy out to a distance of tens of
kiloparsecs (see Figure~\ref{fig:data}).  The stream is oriented such
that it points almost directly at two of M31's satellites, M32 and
NGC205, raising the intriguing possibility that it may be associated
with one or other of these galaxies [although, as \citet{ferg02} point
out, a tidal association with NGC205 is difficult to reconcile with
the absence of any extension in the stream beyond this galaxy].

A subsequent study by \citet{mcc03} confirmed the existence of this
``Southern Stream,'' traced it to larger distances, and was even able to
measure a distance gradient along it, placing it at some 60 degrees to
the line of sight.  Combining the new angular extent with the angle to
the line of sight, they found that the stream is more than 100 kpc in
length.

One of the peculiarities of the stream is that although it is plainly
detected to the south-east of the galaxy, it does not appear in
anything like a symmetric form to the north-west.  There is some
indication of a detection of the stream on this side of the galaxy in
the detailed analysis by \citet{mcc03}, but it is clear that if the
continuation of the stream exists at a similar strength then it must
be oriented in such a way as to be mostly hidden.  It was this
asymmetry that led \citet{ferg02} to hypothesize that the stream's
continuation might turn closer to the disk of M31, and its
re-emergence might be associated with a feature known as the Northern
Spur (see Figure~\ref{fig:data}).

The Northern Spur is a peculiar low-surface-brightness structure
sticking out of M31's disk, which contains a metal-rich stellar
population.  Since it lies in the direction of M31's gaseous warp
\citep{ne77}, its projection away from the plane is usually attributed
to a severe warp in the stellar disk.  However, if so then it would
have to be more extreme than the warps found in any other stellar
disks \citep{wk88}.  It would also have to be an uncomfortably
short-lived asymmetric feature, since there is no comparable spur to
the south of M31.  It therefore seems at least as plausible that this
feature is associated with the star stream rather than the warp
(although there is also always the third possibility that it is
unrelated to either of these phenomena).

A direct link between the Southern Stream and the Northern Spur would
be hard to detect photometrically against the bright disk, especially
since the width of the Southern Stream ($\sim 0.5$ degrees;
\citealt{mcc03}) is comparable to the scale-length of the disk against
which it appears projected, so no sharp features would stand out.
However the kinematics of any stars forming such a link would be
expected to be quite different from those of disk stars, so would be
worth trying to detect.

\section{The kinematics of the stream}\label{sec:kin}

As part of a project to study the stellar kinematics of the disk of
M31, we have obtained radial velocities of 2611 M31 planetary nebulae
(PNe) using a novel purpose-built device, the Planetary Nebula
Spectrograph (PN.S). The instrument, mounted on the William Herschel
Telescope, obtains both positions and line-of-sight velocities for PNe
in a single observation using a form of slit-less spectroscopy;
details of the method and design of the PN.S can be found in
\cite{pns02}.  PNe provide an ideal kinematic tracer of the stellar
population.  They can be readily identified as point-like emission
line sources, and their kinematics can be measured using the same
emission lines.  Since PNe are just ordinary stars that we happen to
catch at the ends of their lives, they are fairly representative of
the bulk stellar population of the galaxy.  Such discrete tracers are
particularly good for searching for stellar streams because they
provide a direct measure of the line-of-sight velocity distribution.
More traditional absorption-line studies of unresolved stellar
components require a deconvolution to obtain a measure of the velocity
distribution, and the noise-amplifying properties of this process
would hide any small kinematic subcomponent like a star stream.  The
resulting PNe data set is presented in Figure~\ref{fig:data}.  The
overall rotation of the disk is clearly visible, and a few of M31's
satellites are also apparent both spatially and from their distinct
kinematics.

To identify any possible stream stars, we need a mechanism for
identifying PNe whose kinematics are inconsistent with their being
members of either M31's bulk disk population or one of the known
satellites.  To this end, we have implemented a simple ``friendless''
algorithm.  For each PN we identify its $N$ nearest neighbours on the
sky, and calculate their mean velocity $\overline{v}$ and velocity
dispersion $\sigma$.  If the PN in question has a velocity that lies
more than $n \times \sigma$ from $\overline{v}$, then it is flagged as
friendless.  This non-parametric approach selects pretty much the same
PNe that one would pick out by eye as having discrepant kinematics,
and turns out to be fairly robust in that the exact values of $N$ and
$n$ chosen do not affect the results dramatically. Simulations of
simple disk models show that only a very small number of ``false
friendless'' PNe would be found in a system that only contains a disk
population.  For the present analysis, we have adopted $N = 30$ and $n
= 4$; the 23 friendless PNe found with these parameters are highlighted
in Figure~\ref{fig:data}.

The first thing that is apparent from the friendless PNe is the
asymmetry in their distribution: away from the minor axis, there are
13 PNe at $X < -0.2$ and $v_{\rm los} < 0$, whereas the corresponding
quadrant on the other side of the galaxy ($X > 0.2$ and $v_{\rm los} >
0$) contains only 5 PNe.  A binomial test reveals that this
lopsidedness is inconsistent with a symmetric distribution at $95\%$
confidence.  Such an asymmetric distribution cannot be reconciled with
any simple explanation for these friendless PNe, such as the extreme
tail of a hot disk population (including a possible thick disk
component), or a relaxed halo population; indeed, any contamination
from these axisymmetric populations would tend to eradicate such an
asymmetry.  The friendless PNe do, however, lie in exactly the region
one would expect to link the Southern Stream and the Northern Spur, so
perhaps they could form the linking part of the stream, or possibly
they could have been kicked out of the original disk through
gravitational interaction with such a stream.  As noted above, the
width of the stream means that no subtler evidence for the stream can
be found in the spatial distribution of friendless PNe, but, as we
shall see in the next section, there is more information to be gleaned
from their kinematics.

\section{The Orbit of the stream}\label{sec:orbit}

We now seek to model the stream features, and test whether the
resulting kinematics are consistent with those of the putative stream
PNe that lie between the photometric features.  A full simulation of
the merger, incorporating realistic tidal stripping and dynamical
friction, is beyond the scope of this report.  However, a simple
single orbit model will go a long way toward describing the shape of
this feature and picking out its likely kinematic signature.

One immediate constraint on possible orbits comes from the rather
sharp angle through which material must turn in order to get from the
Southern Stream to the Northern Spur.  The only way for the stream to
make such a sudden detour is if it follows a fairly radial orbit that
takes it close to the centre of the potential.  Further, the potential
cannot contain a large ``softening'' core, which would inhibit the
strong gravitational interaction necessary to produce the sharp
deviation.  Fortunately, as we shall see below, the rotation curve of
M31 stays nearly flat all the way to very small radii, so there is no
evidence for a significant core in this galaxy.  We therefore adopt a 
flattened singular isothermal potential, $\Phi(R,z) = {1
\over 2}v_c^2\ln (R^2 + z^2/q^2)$ where $R$ and $z$ are polar
coordinates aligned with the disk plane of M31.  For the amplitude, we use
the upper envelope of the PNe velocities, $v_c = 250\,{\rm km}\,{\rm
s}^{-1}$; for the flattening, we adopt a value of $q = 0.9$, in line
with what is found in other galaxies [for example, \citet{maj03}]
although the value of the flattening turns out not to be an important
factor in this case.  The resulting rotation curve is consistent with
previous studies of M31 \citep[e.g., ][]{k89}.

The orbit must fit two major photometric constraints. In the
coordinate system of Figure~\ref{fig:data}, the Southern Stream is a
rather linear feature which enters at projected $X,Y$ values
$(1.0,-2.0)$, and is inclined to the sky by some 60 degrees, with more
negative values of $Y$ lying further away from us \citep{mcc03}.  The
Northern Spur has a sharp outer edge near $(-2.0,0.6)$. Identifying
this edge with a turning point of the orbit implies that the speeds in
$X$ and $Y$ are very small there. Thus, we are left with two free
parameters for the orbit in the Northern Spur, which are the unknown
values of $Z$ and $v_Z$ in this region.  We have therefore searched
the space afforded by these parameters to find if there are any orbits
that reproduce the structure of both the Northern Spur and the
Southern Stream.

\begin{figure*}
\includegraphics[width=177mm]{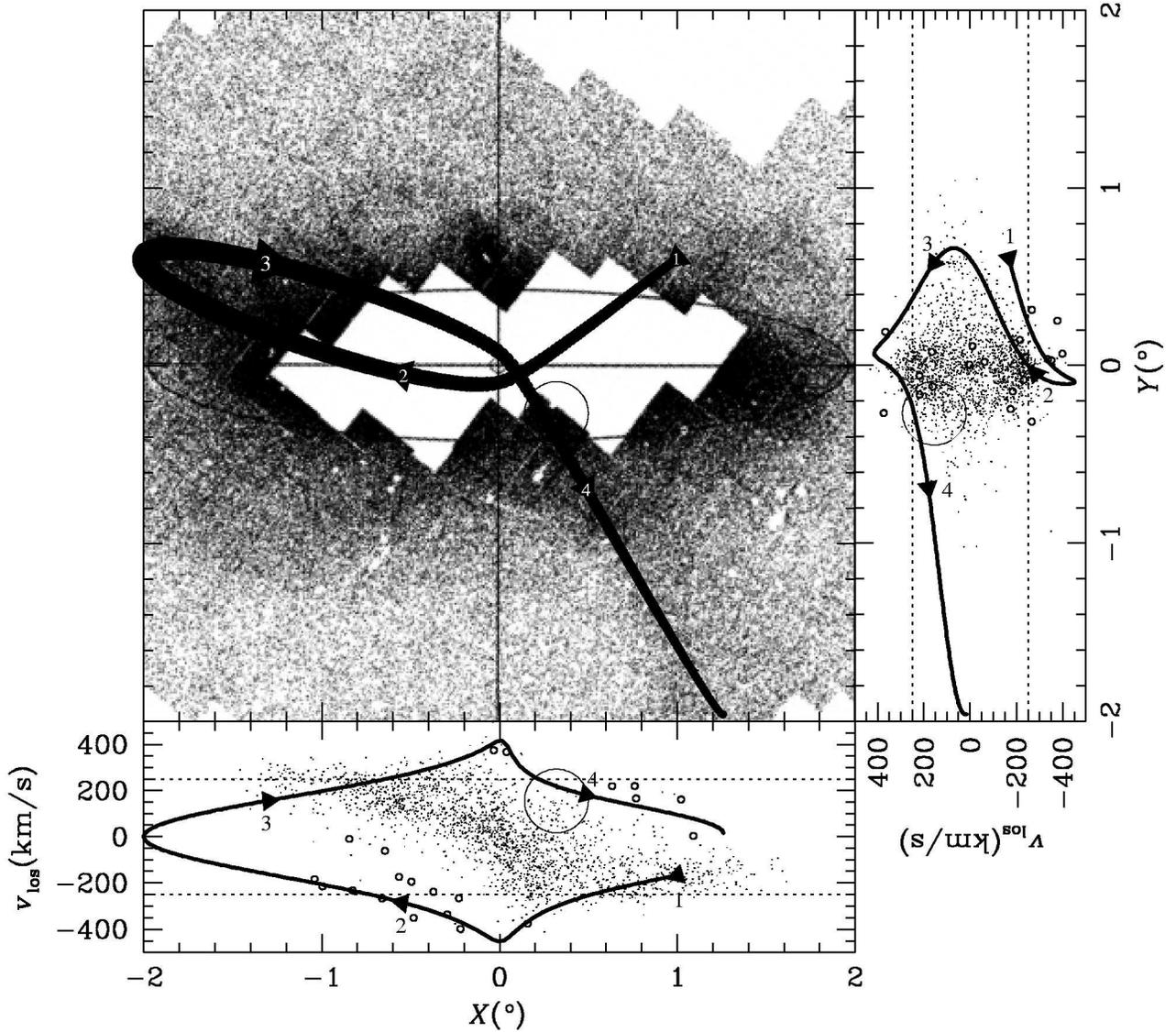}
\caption{An orbit model fitted to the photometric data.  The upper
panel traces the shape of the orbit projected on the photometric data,
with the thickness of the line indicating its distance along the line
of sight, and the numbered arrows showing the suggested direction of
motion along the stream.  The side panels show the projection of this
orbit in line-of-sight velocity with respect to M31 versus distance
along the principal axes, superimposed on the PNe data, with the
friendless PNe highlighted.  These panels show 21 of the friendless
PNe; the remaining two are at $v_{\rm los} < -500\,{\rm km}\,{\rm
s}^{-1}$, so are presumably not associated with M31.  The dotted lines
at $\pm250\,{\rm km}\,{\rm s}^{-1}$ show the adopted circular speed.
The location of M32 is shown as a circle in each panel.  }
\label{fig:model}
\end{figure*}

It turns out that these constraints are strongly restrictive, but an
orbit does exist that meets all these requirements; it is illustrated
in Figure~\ref{fig:model}.  As the upper panel of this figure shows,
in addition to matching the three-dimensional structure of the
Southern Stream and the turning point in the Northern Spur, it is
interesting to note that this orbit originates in the rather confused
``G1'' region [although \citet{ferg02} point out that the colour of
the G1 region differs from that of the stream, bringing into question
any direct association].

The lower panel of Figure~\ref{fig:model} shows the kinematics of the
PNe along the major axis.  Note how the upper envelope of the main
disk population stays constant at $\sim 250\,{\rm km}\,{\rm s}^{-1}$
at all radii, justifying the choice and normalization of the singular
isothermal potential.  This normalization is the only use made of the
kinematic data in the orbit fitting, yet the friendless PNe trace the
projection of the orbit remarkably well.  Although a few of the PNe
closest to the low-velocity envelope of the main disk population are
probably just the tail of the disk component, most follow
the velocity of the orbit, including its decline with radius, rather
closely.  A full model of the stream would have to incorporate the
fact that the stream is not in reality a single orbit, but a family of
adjacent orbits.  The impact of this dispersion is amplified by the
singular nature of the potential, which can scatter adjacent orbits in
significantly different directions, increasing the spread in
velocities: an initial velocity dispersion of $\sim 15\,{\rm km}\,{\rm
s}^{-1}$ in the stream stars can lead to a spread of $\sim 100\,{\rm
km}\,{\rm s}^{-1}$ in the observed line-of-sight velocities after
pericentre passage, much as seen in the data.  

Fortunately, the observed major-axis kinematics of the stream are a
rather generic indicator of any of the orbits that may connect the two
photometric features, irrespective of the details of the adopted
model.  The geometry of the Southern Stream places the stars within it
on an almost radial orbit, so that when they arrive in the disk of M31
they will be travelling at high velocities.  However, the sharp edge
of the Northern Spur is the signature of a turning point of the orbit,
indicative of low velocities.  Thus, stars on this link would be
expected to show a velocity gradient along M31's major axis from above
the circular speed near the Southern Stream to close to zero at the
radius of the Northern Spur.  The width of the stream and the
concentration of the survey toward the plane of M31 means that there
is less information in the minor axis projection.  However, there is a
notable concentration of PNe around arrow 2 on the stream orbit in
both velocity projections in Figure~\ref{fig:model}.  Thus, both the
asymmetry in the distribution of friendless PNe and their major axis
kinematics point to them providing the ``missing link'' between the
Southern Stream and the Northern Spur.

One interesting question is whether the progenitor of the stream still
exists as a coherent object, or whether the detritus is all that
remains.  It is interesting to note that the orbit passes very close
to M32 both in position [as pointed out by \citet{ferg02}] and in
velocity.  If this is not just a chance superposition, then it implies
that the streams are the tidal tails of M32 that have been ripped off
as it orbits M31.  Several less direct lines of argument support this
hypothesis:
\begin{enumerate}
\item M32, the Northern spur and the Southern stream all have a red
  giant branch that is particularly red compared to the rest of the halo
  of M31 \citep{ferg02}.
\item M32 sits more or less in the middle of the populated part of the
  orbit: the Southern Stream (which leads M32 in the model) is seen to
  extend to at least 3 degrees towards the south, comparable in length
  to the trailing part of the orbit which extends through the Northern
  Spur and back towards the center of M31.
\item M32 has the appearance of a highly tidally stripped galaxy.
  Crudely speaking, one might expect such a tidally stripped galaxy to
  have lost around half its total luminosity: much less than half, and
  it would not appear to have been stripped; much more than half and it
  would have disappeared entirely.  It is therefore interesting to
  note that the number of PNe detected in the stream is similar to
  what is detected in M32 itself.
\end{enumerate}
Note that this model implies that M32 at present lies about 4~kpc
behind the centre of M31, but in front of the disk plane.  Whether M32
lies behind or in front of M31 is currently an open question
\citep{mat98}, with some indications that it may lie in front
\citep{fjj78}; a better relative distance measure would provide a
further important check on this model.

It is also worth noting that the direction of motion along the stream
is not, as yet, very tightly constrained.  If the stars were
travelling in the opposite direction, from the Southern Stream toward
the Northern Spur, then the only difference in Figure~\ref{fig:model}
would be that the orbit in the lower panel would be reflected about
$v_{\rm los} = 0$.  Since the shape of the orbit in this projection is
approximately symmetric about $v_{\rm los} = 0$, the difference in the
quality of fit is rather slight, so this possibility cannot be ruled
out.  One significant difference in this configuration is that M32 is
no longer simultaneously coincident with the stream both spatially and
kinematically, so it is still possible that the stream is the remnant
of another satellite altogether.  The recently discovered satellite
And~VIII which \cite{mor03} identified from a concentration of PNe,
faint HI clouds, and globular clusters with radial velocity with
respect to M31 near $-204\rm km\,s^{-1}$ would be a candidate: this
tidally stretched satellite occupies the region $X=0-1$, $Y\sim-0.5$
(just beyond the edge of the PN.S survey), near M32 and the point
where the Southern Stream meets the disk of M31.

\section{Conclusions}\label{sec:conc}

We have demonstrated that the star stream to the south of M31 and the
spur to the north of the system can be modeled as parts of a single
coherent merger stream stretching over some hundreds of kiloparsecs.
As well as explaining where the Southern Stream goes after
disappearing into M31, this model also eliminates the awkward need to
invoke the most extreme stellar warp known as an explanation for the
Northern Spur.  This is not to say that the stream is unrelated to the
gaseous warp, however, as such a merger event could well play a role
in exciting the warp \citep{qhf93}.

The model also makes a prediction as to where one might be able to
detect the stream kinematically, and observations of the PNe in the
disk of M31 show that there are stars at exactly the velocities that
one would expect.  These data trace the stream in the bright disk
region where there is no chance of detecting it photometrically, and
provide a direct link between the two photometric features at larger
radii.

It is at present unclear whether the satellite whose debris makes up
the Stream is still present or not. Both M32 and And~VIII are possible
parents, though they imply different orbit solutions. Direct
measurements of the radial velocity of the Southern Stream will be
required to distinguish between the possibilities.

\section*{Acknowledgments}
This research is based on data obtained using the William Herschel
Telescope operated by the Isaac Newton Group in La Palma; the support
and advice of the ING staff is gratefully acknowledged.  We would also
like to thank the referee for many helpful comments on the
manuscript.

\label{lastpage}

\end{document}